\documentclass[12pt]{nostro}

\usepackage{epsfig,amssymb,amsfonts,wasysym}\parskip 5pt plus 1pt
\newcommand{\be}{\begin{equation}}
\newcommand{\ee}{\end{equation}}
\newcommand{\bea}{\begin{eqnarray}}
\newcommand{\eea}{\end{eqnarray}}

\def\simge{\mathrel{%
   \rlap{\raise 0.511ex \hbox{$>$}}{\lower 0.511ex \hbox{$\sim$}}}}
\def\simle{\mathrel{
   \rlap{\raise 0.511ex \hbox{$<$}}{\lower 0.511ex \hbox{$\sim$}}}}

\begin{document}
%
\begin{frontmatter}
\thispagestyle{empty}
\begin{flushright}
\end{flushright}

\title{A new generation photodetector for astroparticle physics: the VSiPMT}
Authors
\author[INFN,UNI]{G. Barbarino}
\author[INFN,UNI]{F.C.T.Barbato}
\author[UNI]{L. Campajola}
\author[INFN,UNI]{F. Canfora}
\author[INFN]{R. de Asmundis}
\author[INFN,UNI]{G. De Rosa}
\author[INFN,UNI]{G. Fiorillo}
\author[INFN]{P. Migliozzi}
\author[INFN]{C.M. Mollo}
\author[INFN]{B. Rossi}
\author[INFN,UNI]{D. Vivolo}


\address[INFN]{Istituto Nazionale di Fisica Nucleare - Sezione di Napoli, Complesso di Monte S. Angelo Edificio 6, via Cintia 80126 Napoli, Italy}
\address[UNI]{Universit\`{a} degli Studi di Napoli "Federico II", Dipartimento di Fisica, via Cintia 80126 Napoli, Italy}

\vspace{.3cm}

\begin{abstract}
The VSiPMT (Vacuum Silicon PhotoMultiplier Tube) is an innovative design we
proposed for a revolutionary photon detector. The main idea is to replace
the classical dynode chain of a PMT with a SiPM (G-APD), the latter acting as
an electron detector and amplifier. The aim is to match the large sensitive
area of a photocathode with the performance of the SiPM technology. The
VSiPMT has many attractive features. In particular, a low power consumption and an excellent photon counting capability. 
To prove the feasibility of the idea we first tested the performance of a special
non-windowed SiPM by Hamamatsu (MPPC) as electron detector and current amplifier. Thanks to this result Hamamatsu realized two VSiPMT industrial prototypes.
In this work, we present the results of a full characterization of the VSiPMT prototype.
\end{abstract}

\vspace*{\stretch{2}}
\begin{flushleft}
  \vskip 2cm
  \small
\end{flushleft}
\end{frontmatter}

\section{Introduction}
\begin{figure}[h!]
	\centering
\includegraphics[scale=.4]{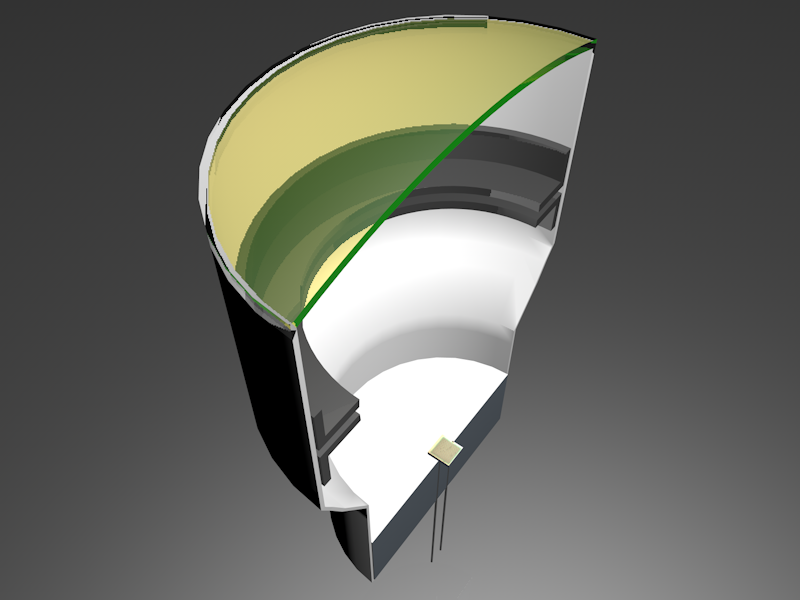} 
		\caption{The VSiPMT conceptual design.}\label{VSiPMTdesign}
	\end{figure}

The first documented photomultiplier demonstration dates to the early 1934 accomplishments of an RCA group based in Harrison, NJ. Harley Iams and Bernard Salzberg were the first to integrate a photoelectric-effect cathode and single secondary emission amplification stage in a single vacuum envelope and the first to characterize its performance as a photomultiplier with electron amplification gain \cite{1934}. 
Currently PMTs are largely used in many experiments. Focusing on the astroparticle physics area, there are three classes of experiments that are very challenging both from their physics goals and the adopted detector technology: Neutrino Telescopes, Cherenkov Telescopes Arrays, Dark Matter searches.\\
Neutrino Telescopes are aiming at the detection of very high-energy neutrinos produced by galactic and/or extra-galactic sources. The detector consists of tens of thousands of photodetectors that are used to instrument cubic kilometers of ice/water located at elevated depth. The photomultipliers used in the ongoing experiments (ANTARES, Km3NeT, ICECUBE), although performing very well, have the following drawbacks \cite{KM3}:
	\begin{itemize}
		\item the power consumption is not negligible, imposing strong requirements on the mechanical structure that 		
		hosts them;
		\item the resolution of photoelectrons is poor;
		\item the Transit Time Spread (TTS) is not better than $1-2\:ns$. 
	\end{itemize}
For these experiments, the TTS is a critical parameter since it strongly affects the track reconstruction. On the other side requiring PMTs with a small TTS increases the costs.\\
Cherenkov Telescopes Array use PMTs to detect light focused in the focal planes of the telescopes. As discussed in \cite{CTA} the most critical characteristics of the PMTs in this field are: pulse shape, gain, temperature dependence, quantum efficiency, noise and aging. Another aspect it is worth to mention is that in order to cover the large area camera, the number of channels is huge. As an example, in the medium size telescope of the CTA experiment the camera will consist of about 11000 pixels, each with physical size about $6\times 6\: mm^2$, with a high dark count rate. \\
To reduce the dark noise down to few MHz, it has been proposed to operate the photomultipliers at moderately low temperatures, around 10 $^\circ$C.\\ 
PMTs play a crucial role also in direct dark matter search experiments with noble liquids \cite{DarkMatter}. In these applications PMTs are used to convert into an electrical signal the low quantity of light emitted in the interaction of dark matter with the detector target. The requirements are the following: 
	\begin{itemize}
		\item ability to work at cryogenic temperature;
		\item single photoelectron detection capability;
		\item high single photoelectron resolution;
		\item radioactivity well below $1\:mBq$. This is a major challenge found in most dark matter detection experiments.
	\end{itemize}
To overcome these limits, we invented a new high-gain, silicon-based photodetector, the Vacuum Silicon Photomultiplier Tube \cite{IdeaIbrido}. The innovative idea is to replace the traditional dynode chain of a PMT with a SiPM (Fig. \ref{VSiPMTdesign}), the latter acting as an electron detector and then as a current amplifier with a gain of $>10^6$, that is similar to the one of standard PMTs.
The realization of the VSiPMT with a photocathode of at least 1 inch diameter will have a tremendous impact on astroparticle physics experiments by starting a new generation of photodetectors which exhibit several attractive features such as: excellent photon counting, high gain ($>10^6$), negligible power consumption($\sim\: nW$), small TTS ($<\: ns$), simplicity, compactness and robustness.\\ 
In this paper we discuss how the VSiPMT deals with the highly demanding requirements of the astro-particle experiments discussed above. 

\section{The VSiPMT prototype}
\begin{figure}[h!]
	\centering
		\includegraphics[scale=.45]{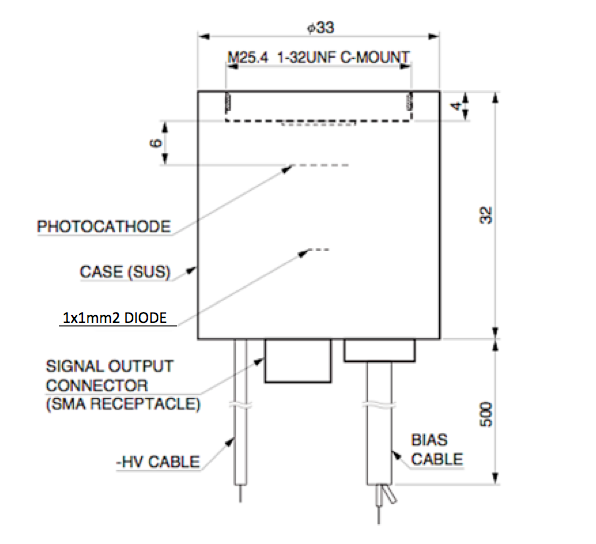} 
		\caption{Scheme of the VSiPMT prototype realized by Hamamatsu.}\label{schema_prototipo}
	\end{figure}

The R\&D to demonstrate the VSiPMT feasibility started with a full Geant-4 simulation and then by testing the response of a special non-windowed MPPC by Hamamatsu to an electron beam. The MPPC showed quite well separated peaks and, consequently, electron counting capabilities providing the first proof of feasibility of the device \cite{simulazioni} \cite{fattibilita} \cite{cern}. These results encouraged Hamamatsu to develop two VSiPMT prototypes (EB-MPPC050 (ZJ5025) and EB-MPPC100 (ZJ4991)), whose characteristics have been carefully studied in the INFN-Napoli laboratories.
	\begin{figure}[h!]
	\centering
		\includegraphics[scale=.6]{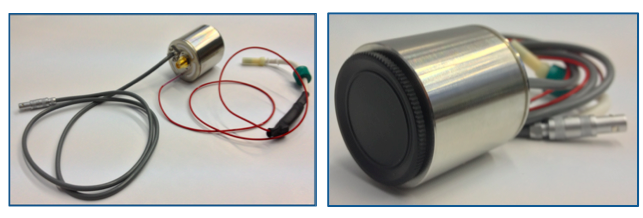} 
		\caption{The VSiPMT prototype by Hamamatsu.}\label{prototipo}
	\end{figure}\\
One of the two prototypes is shown in Fig. \ref{prototipo}. The absence of the divider has several very important consequences:
	\begin{enumerate}
		\item the device is compact;
		\item there are only three connections: two cables to supply power to the photocathode and to the MPPC (the thin cable on the right and the lemo cable on 
		the left, respectively) and one SMA output for signal readout.
\end{enumerate}
The prototypes have the same envelope with a $7\times7\:mm^2$ borosilicate glass entrance window and a GaAsP photocathode with a $3\:mm$ diameter and a spectral response in the range between $300$ and $750 \:nm$ (Fig. \ref{schema_prototipo}).
	\\
The two devices differ only in the characteristics of the MPPCs used, see Table \ref{differenze}, in this paper we will relate to ZJ5025 results.
	\begin{table}[h!] \caption{Table of the VSiPMT characteristics.}\label{differenze}
	\centering	
	\begin{tabular}{rcc}
		\hline \textsc{Prototype} & \textbf{ZJ5025} & \textbf{ZJ4991} \\ 
		\hline \textsc{MPPC Area ($mm^2$)} & $1\times1$ & $1\times1$ \\ 
		 \textsc{Cell Size ($\mu m$)} & 50 & 100 \\ 
		 \textsc{Total Number of Cells} & 400 & 100 \\ 
		 \textsc{Fill Factor} & 61 & 78 \\ 		 
		 \textsc{Optimized Configuration} & $p^+nn^+$ & $p^+nn^+$ \\ 
		 \textsc{MPPC Operation Voltage (V)} & 72.5 & 72.4 \\ 
		 \textsc{Photocathode Power Supply (kV)} & -3.2 & -3.2 \\ 
		\hline 
	\end{tabular}
	\end{table}

\section{The amplifier}
The typical VSiPMT single photoelectron (\textit{spe}) signal amplitude is a few mV, thus an external amplification stage is required.\\
Therefore, we built three custom amplifiers based on the AD8009 OpAmp by Analog Device, a 5 GHz bandwidth Operational Amplifier used in non-inverting configuration. 
The three versions present different gains (10, 15 and 20 V/V, respectively), 50 Ohm input impedance, 50 Ohm output line. \\
For these prototypes we needed to use this amplifier, with $5-10\:mW$ power consumption. However, a VSiPMT made with the next generation of SiPMs will have a higher gain, therefore it will be not necessary to use an external amplifier.
	\begin{figure}[h!]
	\centering
		\includegraphics[scale=.7]{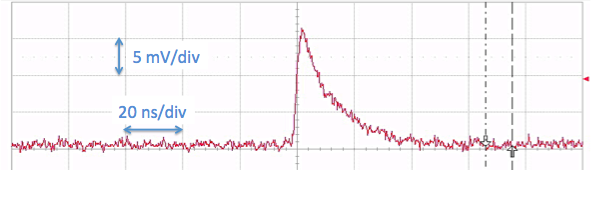} 
		\caption{Tyoical amplified VSiPMT spe signal.}\label{segnaleamplificato}
	\end{figure}\\
The shape of the output signal from the amplifier with gain 20 is shown in Fig. \ref{segnaleamplificato}. The typical amplitude after the amplification turns out to be few tens of mV with rise time of few ns.

\section{The VSiPMT characterization: experimental setup and tests}
The prototype underwent many tests summarized here below and discussed in the following sections, in order to achieve a full characterization of:
	\begin{itemize}
		\item signal quality, stability and photon counting capability;
		\item detection efficiency;
		\item gain;
		\item photocathode scan;
		\item Transit Time Spread;
		\item afterpulses and dark counts;
		\item linearity and dynamic range.
	\end{itemize}
A picosecond pulsed laser emitting in the blue region ($407\:nm$) has been used.
The laser light is sent through a system of optical fibres to the VSiPMT, its intensity can be varied by using differently calibrated filters (down to single photoelectron condition)
and is continuously monitored by a power meter. The amplified signal output is directly sent to the oscilloscope and to a computerised digitizing system (CAEN V1724E, 12 bit, 4 ns sampling rate) which are triggered by the laser.\\ 

\section{Signal quality, stability and photon counting capability}
\begin{figure}
	\centering
		\includegraphics[scale=.3]{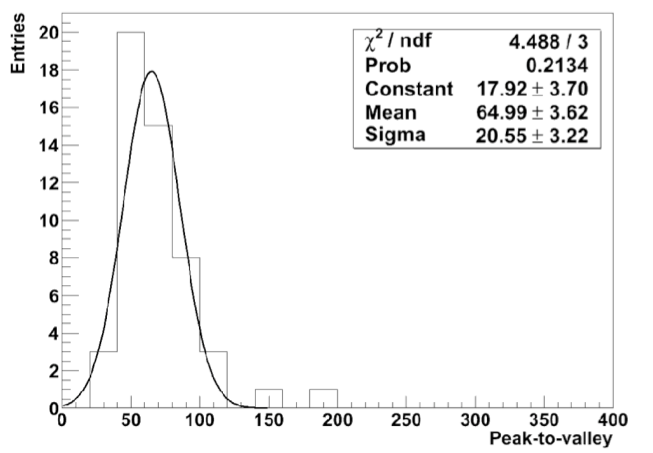}
		\caption{Plot of the VSiPMT peak to valley ratio.}\label{PV} 
	\end{figure}

\begin{figure}[h!]
	\centering
		\includegraphics[scale=.35]{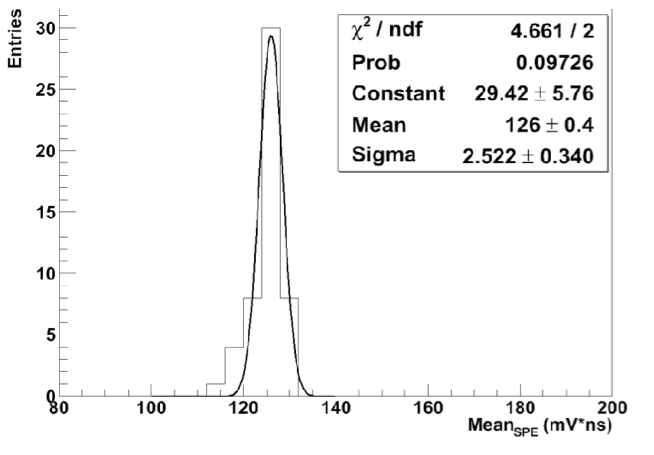} 
		\caption{VSiPMT gain stability in time.}\label{stability}
	\end{figure}

First of all the VSiPMT signal has been studied by illuminating the photocathode with a low intensity laser beam. 
	\begin{figure}[h!]
	\centering
		\includegraphics[scale=.45]{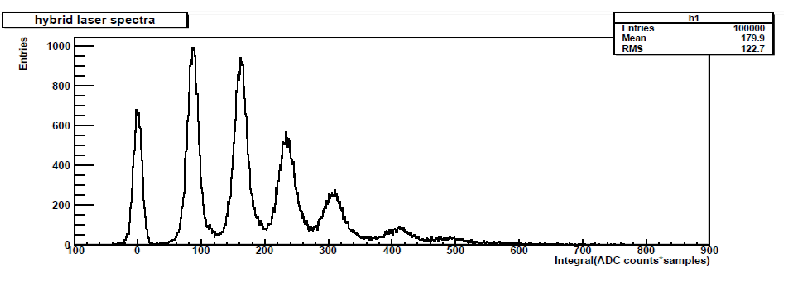} 
		\caption{A typical VSiPMT spectrum.}\label{segnale_oscilloscopio}
	\end{figure}\\
A typical charge spectrum is shown in Fig. \ref{segnale_oscilloscopio}, the remarkable separation between the peaks demonstrates the excellent photon counting capability of such a device. 
The peak-to-valley ratio has been measured to be about 60 in the single photon condition, Fig. \ref{PV}.
The stability of the signal has been proved by acquiring 100000 waveforms, with low intensity laser light, every $20\:min$ for $20\:hours$. The device exhibits an excellent gain stability (better than 2\%) in time, see  Fig. \ref{stability}.
	\\

\section{Detection efficiency: stability and uniformity}
\begin{figure}[h!]
	\centering
		\includegraphics[scale=.45]{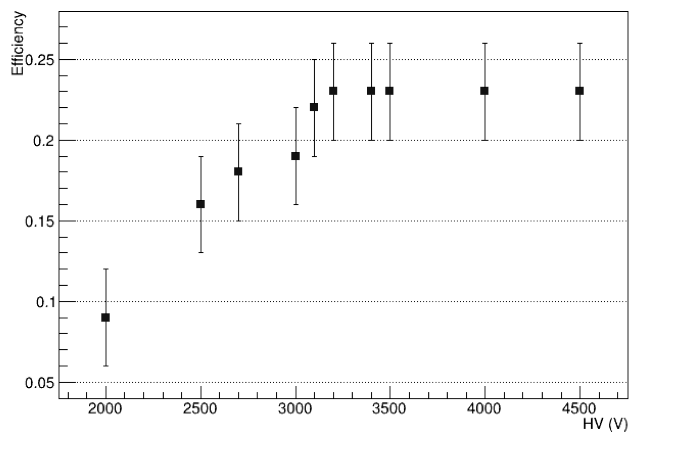} 
		\caption{Results of the photon detection efficiency test.}\label{OperatingPoint}
	\end{figure}

\begin{figure}[h!]
	\centering
		\includegraphics[scale=.37]{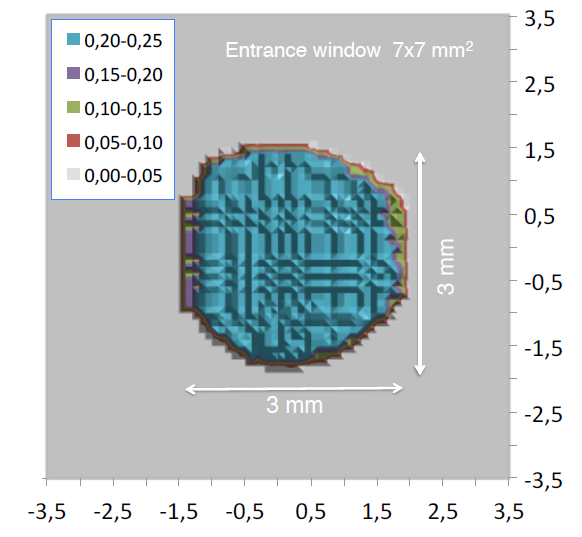} 
		\caption{Results of the x-y scan of the photocathode of the prototype.}\label{scan}
	\end{figure}

The photon detection efficiency (\textit{PDE}) of the device can be factorized as follows:
	\begin{equation}
		PDE_{VSiPMT} =QE\cdot \epsilon_{MPPC}\cdot \epsilon_{coll}
	\end{equation}
where QE is the photocathode quantum efficiency, $\epsilon_{MPPC}$ is the MPPC electron detection efficiency and $\epsilon_{coll}$ is the collection efficiency of the photoelectrons on the SiPM surface. $\epsilon_{MPPC}$ is equal to the fill factor, since electrons penetrating the MPPC are always detected.\\
First and foremost we evaluated the PDE.
The device has been illuminated with few photons per pulse by using an optical fibre, during the test the laser beam intensity has been monitored by a power meter. The PDE is defined as:
	\begin{equation}
		PDE = \frac{N_{pe}}{N_{ph}}		
	\end{equation}

where $N_{pe}$ is the number of fired cells and $N_{ph}$ is the number of photons per laser pulse hitting the photocathode.\\
In the same light conditions, we tested the PDE as a function of HV, by varying the photocathode power supply in steps of 100 volts. The PDE behaviour has been measured, see Fig. \ref{OperatingPoint}.
The PDE of the device becomes stable around $-3\: kV$, with a plateau till $-5\: kV$. The operating point for the high-voltage supply has been fixed to $-3.2\: kV$, corresponding to $\sim 23$\% PDE value. 
In such a device, the plateau region is linked to the energy threshold of the photoelectrons. The photoelectrons need a minimum energy to enter into the silicon bulk and, consequently, to produce a signal. Therefore, the high voltage is required only to drive photoelectrons to the SiPM surface and to give them the right energy. Thus, differently from a classical PMT, once the device is in the correct plateau, there is no need for high voltage stabilization.\\
We also performed an x-y scan of the photocathode, to probe the uniformity on the entire sensitive surface by estimating the local PDE. For this measurement, we used a micrometric x-y motorized pantograph.
Fig. \ref{scan} shows the result of the x-y scan of the $3 mm \: \varnothing$ GaAsP photocathode in the $7 \times 7 \:mm^2$ entrance window.
The measured PDE is ranging between 20\% an 25\%, see Fig. \ref{scan}, showing a good uniformity through the whole photocathode surface. The boundary effects are negligible.

\section{Gain}
In a VSiPMT the gain is obtained by the electrons crossing the Geiger region of the SiPM. Thus, a standard current signal is given for each fired cell. On the basis of this working principle, we obtained a value of the gain as the ratio between the single pixel charge output signal and the electron charge.\\
For this measurement, the output signal needs to be converted in the corresponding equivalent charge and the contribution of the external amplifier must be taken into account.\\
A measurement of the gain value with respect to the voltage supply has been carried out at $ -2.2\: kV$ (in the ramp up region) and at $-3.2\: kV$ (in the plateau region) and keeping the MPPC bias voltage at $72.5\: V$. As expected we found that the value for the gain at $72.5\:V$ is about $5.6 \cdot 10^5$, that is the expected value for the MPPC. Then, we performed a further measurement by setting the HV to $-3.2\: kV$ and varying the MPPC bias voltage from $72.1 \:V$ to $73.0\: V$ in steps of $0.1\: V$.
	\begin{figure}[h!]
	\centering
		\includegraphics[scale=.45]{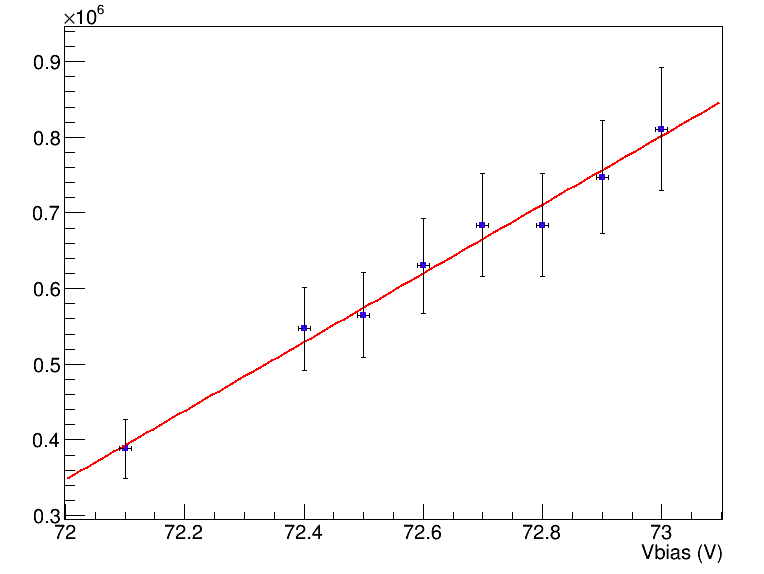} 
		\caption{VSiPMT gain workfunction.}\label{GainTrend}
	\end{figure} 
Fig. \ref{GainTrend} shows the linear trend of the gain as a function of the MPPC power supply, we estimate a systematic error of about 10\%.

\section{Transit Time Spread}
\begin{figure}[h!]
	\centering
		\includegraphics[scale=.35]{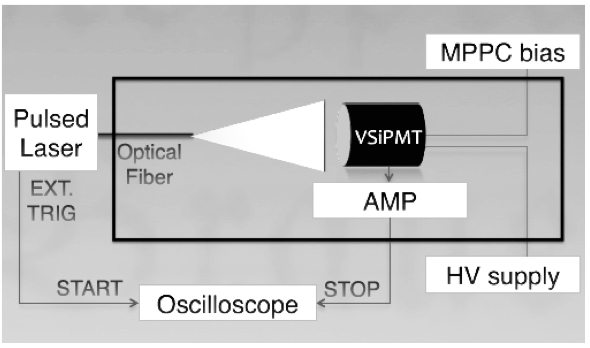} 
		\caption{Experimental setup scheme for the TTS measurement.}\label{TTSscheme}
	\end{figure}

\begin{figure}[h!]
		\centering
		\includegraphics[scale=.65]{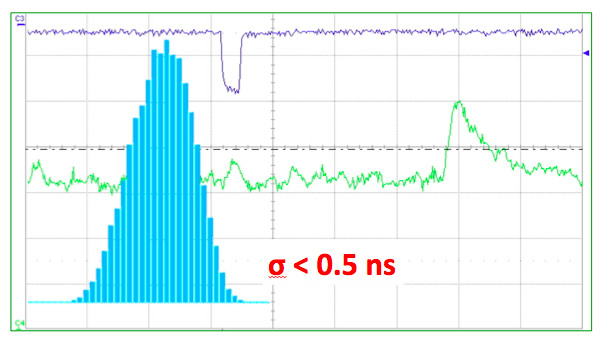} 
		\caption{Printscreen of the oscilloscope. The blue squarewave is the trigger signal from the oscilloscope. The green waveform is the VSiPMT spe signal. The histogram is the distribution of the arrival time of the VSiPMT signal with respect to the laser trigger.}\label{tts}
	\end{figure}

The timing plays a crucial role in those experiments, like water Cherenkov neutrino telescopes, where particle direction has to be reconstructed with high accuracy. A short TTS is mandatory to disentangle different events and to reconstruct the particle direction with a good angular resolution.\\
We expect the TTS to be smaller for the VSiPMT than for a standard PMT. For a standard PMT, indeed, there are two contributions to the TTS: one is due to the different trajectories of the photoelectrons going from the photocathode to the first dynode, the other is due to the different trajectories of the seconday electrons between the dynode chain.\\
In the VSiPMT the dynode chain is replaced by a SiPM, thus for such a device the TTS is simply due to the photoelectron trajectories between the photocathode and the SiPM.\\
The TTS measurement has been performed by using the experimental setup shown in Fig. \ref{TTSscheme}.
The sync pulse of the laser is used as trigger, while the output signal of the VSiPMT is fed as stop signal via a discriminator. The time difference between the start and the stop signal has been measured. Taking into account an eventual laser jitter, we estimated the TTS upper limit, as the $\sigma$ of the distribution, to be $<0.5\: ns$, see Fig. \ref{tts}.

\section{Dark count rate}
The MPPC dark count rate is known to be strongly dependent on the bias voltage and on the pe threshold. For this reason, measurements of the dark count rate at different bias voltage and different thresholds (0.5, 1.5 and 2.5 pe respectively) have been done.\\
	\begin{figure}[h!]
	\centering
		\includegraphics[scale=.45]{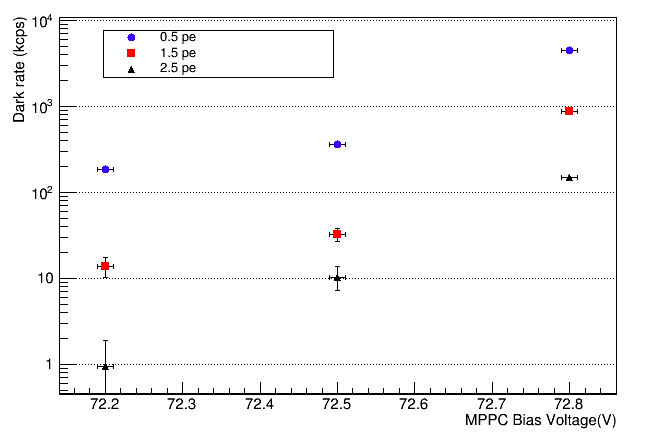} 
		\caption{Dark count rate of the prototypes at 0.5, 1.5 and 2.5 pe value threshold.}\label{DarkCount}
	\end{figure} 
Fig. \ref{DarkCount} shows the results of the dark count rate of the prototype.\\ 
Moreover, to prove that the VSiPMT dark count rate depends basically on the MPPC used, we fixed the bias voltage and we performed the measurements in two different configurations: with HV set to the operating point and with no photocathode power supply, too. \\
	\begin{table}[h!]\caption{VSiPMT dark counts at $V_{bias} = 72.5\:V$.}\label{dc_onoff}
	\centering
		\begin{tabular}{ccc}
			\hline Threshold (pe) & $Rate_{DC}$ (kcounts/s)  & HV \\ 
			 \hline 0.5 & $348\pm 0.6$ & ON \\ 
			 0.5 & $361\pm 0.6$ & OFF \\ 
			\hline 1.5 & $35\pm 0.2$ & ON \\ 
			 1.5 & $33\pm 0.2$ & OFF \\ 
			\hline 2.5 & $10\pm 0.1$ & ON \\ 
			 2.5 & $11\pm 0.1$ & OFF \\ 
			\hline 
		\end{tabular} 
	\end{table}
No differences have been found between the two configurations (see Table \ref{dc_onoff}), thus demonstrating that the dark count rate depends only on the MPPC.

\section{Afterpulses}
\begin{figure}[h!]
	\centering
			\includegraphics[scale=.45]{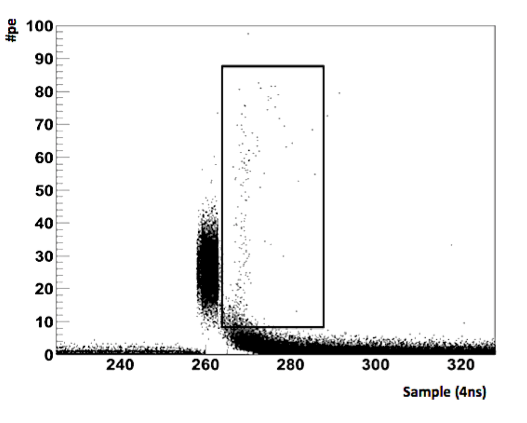} 
		\caption{Signal amplitude distribution vs arrival time of the device.}\label{afterpulse}
	\end{figure}

The VSiPMT suffers of two different afterpulse classes: the SiPM internal afterpulses and the vacuum tube afterpulses.\\
The former are known to be due to silicon impurities and are characterized by a small amplitude signal (up to 3 pe) and higher frequency with respect to the latter: their contribution has been measured to be less than 10\%. Vacuum afterpulses are characterized by a high amplitude signal (up to 80 pe) and a very low occurrence frequency. We expect that this kind of afterpulses are generated by the interaction of electrons with the residual gases or degased materials present in the tube after evacuation. In this case the arrival time of the afterpulse depends on the ion mass and on the HV applied between the photocathode and the MPPC, while the afterpulse amplitude depends only on the HV \cite{RCA}. Thus, we measured the afterpulses arrival time and their amplitude at different HV, see Tab. \ref{afterpulseAT}. These measurements, together with the afterpulses arrival time distribution (Fig. \ref{afterpulse}), indicate they are due to the ionization of residual gases as expected. The residual gas contribution (points in the rectangle in Fig. \ref{afterpulse}) is very low $< 0.02$\% .

\begin{table}[h!]\caption{Measured values for afterpulses arrival time and amplitude.}\label{afterpulseAT}
\centering
	\begin{tabular}{|c|c|c|}
		\hline HV(kV) & Delay(ns)  & Amplitude ($N_{pe}$) \\ 
		\hline  2 & 52.8 & 10-25 \\ 
		\hline  3 & 43.6 & 18-60 \\ 
		\hline  4 & 38.4 & 22-80 \\ 
		\hline 
	\end{tabular} 
\end{table}

\section{Linearity and dynamic range}
\begin{figure}[h!]
\centering
\includegraphics[scale=.35]{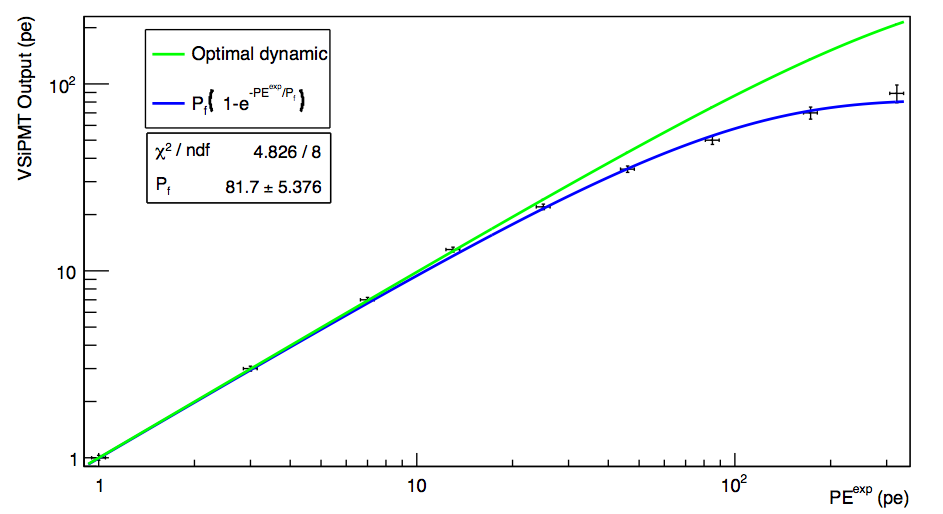} 
\caption{Dynamic range of the prototype: experimental points, fit curve (blue) and  optimal dynamic curve (green).}\label{dyn_50um}
\end{figure}

The dynamic range of a VSiPMT depends on the total number of cells of the SiPM and on the capability of focusing uniformly photoelectrons over the SiPM surface.
\\Given the total number of available cells ($N_{cells}$) of the SiPM, the number of fired cells ($N_{pe}$) as a function of the number of incident photons ($N_{ph}$) and of the PDE is given by:
\begin{equation}
\label{dyn_vsipmt}
N_{pe} (N_{cells}, \lambda, V)= N_{cells} \times \left[ 1- \exp \left( \frac{-N_{ph} \times PDE(\lambda, V)}{N_{cells}} \right) \right].
\end{equation}
The exponential trend of the SiPM dynamic range sets a strict constraint on the performance of the VSiPMT. Indeed, the Eq. \ref{dyn_vsipmt} represents the ideal maximum dynamics for the prototype under test, with $N_{cells}=400$.
\\Another major contribution to the dynamic range of the VSiPMT is provided by the focusing. The GaAsP photocathode of the VSiPMT prototype has a circular shape. Therefore, taking into account the axial symmetry of the electric field generated by the focusing ring, an approximately circular photoelectron spot is reasonably expected. This means that an optimal focusing condition is achieved if the photoelectron beam spot is perfectly inscribed in the SiPM square target. In this case, no more than the $85\%$ of the total number of cells of the SiPM can be fired, therefore the optimized dynamic range for the prototype is that plotted in Fig. \ref{dyn_50um} (green curve), corresponding to $N_{cells}=340$.
\\We have performed a measurement of the dynamic range of the prototype. As a first step, the prototype has been kept in single photoelectron condition attenuating the incident laser light with an appropriate combination of the grey filters. Then, the number of the expected photoelectrons ($PE^{exp}$) has been gradually increased reducing the attenuation. At each step, the VSiPMT output has been measured (number of fired cells, $N_{pe}$). The experimental points are plotted in Fig. \ref{dyn_50um}. The measured values are distributed accordingly to the optimal dynamic range curve up to $\sim 20\:pe$. In correspondence to this value, a saturation effect starts to appear, with the following points increasingly deviating from the theoretical curve. This represents a robust hint for a too strong focusing. If such condition holds, the effective number of fired SiPM cells ($N_{eff}$) is smaller with respect to the optimal case, therefore the VSiPMT dynamic range curve is given by Eq. \ref{dyn_vsipmt} with $N_{eff}$ replacing $N_{cells}$. 
\\$N_{eff}$ has been estimated fitting the measured points to the theoretical curve of Eq. \ref{dyn_vsipmt}
The resulting curve, plotted in Fig. \ref {dyn_50um} (blue), indicates a number of effective cells of $\sim 80$, corresponding to a photoelectron beam spot size of $\approx 0.4\:mm$. Such value is in fair agreement with the indications provided by the manufacturer.
\\The prototype under test  has been realized with the primary aim of testing the working principle of the device and hence its feasibility. No dedicated study has been performed about focusing, therefore the adopted solution is not optimized, thus explaining the reduced dynamic range with respect to the optimal case. 
\\It is worth to notice that there is plenty of room to improve the dynamic range in a second generation of VSiPMT prototypes, with the adoption of a SiPM with a larger number of pixels (striking a balance with the fill factor) and with the realization of an optimized focusing for the maximization of the effective number of fired SiPM cells.

\section{Conclusion}
VSiPMT prototypes show excellent performances: unrivalled photon counting capability, low power consumption (thanks to the absence of the voltage divider), a very easy stabilization and a small TTS ($< 0.5\: ns$). All these features are due to a new high gain concept, thus not implying any additional cost.\\
On the other side, these devices also suffer of some drawbacks as high dark count rate and small linear range, the latter related to the number of pixels of the MPPC. Both problems are being studied. The former can be partially solved by using the new generation of MPPC by Hamamatsu, already available for sale, for which the dark count rate is reduced by a factor 10. The dynamic range of this device strongly depends on the number of pixels of the SiPM used, the greater is the number of pixels the wider is the dynamic range. For these reasons, a SiPM with new optimized shape and a new focusing technique is currently under study.\\
Although R\&D work is still needed to assess the ultimate features of the proposed device, the results presented in this paper unambiguously prove the feasibility of the new VSiPMT photodetector. Furthermore, it has been demonstrated that the VSiPMT has the potentiality to fulfill the requirements of the next generation of astroparticle physics experiments.

\section*{Acknowledgements}
The authors wish to thank Hamamatsu Photonics  K. K. for the realization of the prototypes and for the support during the characterization phase, and Mr Mario Borriello for the constant help and for the precious technical support during setup and testing activities.

\end{document}